\documentclass[preprint,showpacs,groupedaddress,amsmath,amssymb]{revtex4}
\usepackage{graphicx}% Include figure files
\usepackage{dcolumn}% Align table columns on decimal point
\usepackage{multirow}
\usepackage{amsfonts}

\newcommand{\beq}{\begin{equation}}
\newcommand{\eeq}{\end{equation}}
\newcommand{\bea}{\begin{align}}
\newcommand{\eea}{\end{align}}
\newcommand{\epsi}{\varepsilon}
\newcommand{\Cth}{\cos \theta}
\newcommand{\Sth}{\sin \theta}
\newcommand{\del}{\nabla}

\newcommand{\rhat}{\hat{r}}

\newcommand{\phihat}{\hat{\phi}}

\newcommand{\mbf}[1]{\mathbf{#1}}

\newcommand{\Bpol}{B_\theta}
\newcommand{\Btor}{B_\phi}
\newcommand{\divr}{\nabla \cdot}
\newcommand{\curl}{\nabla \times}
\newcommand{\ddrp}{\partial p / \partial r}
\newcommand{\bhat}{\hat{b}}
\newcommand{\para}{\parallel}

\makeatletter
\def\Tbar{\mathchoice
   {\TTbar\displaystyle\textstyle{-}}%
   {\TTbar\textstyle\scriptstyle{-}}%
   {\TTbar\scriptstyle\scriptscriptstyle{-}}%
   {\TTbar\scriptscriptstyle\scriptscriptstyle{-}}%
   \!T}
\def\TTbar#1#2#3{{\setbox0=\hbox{$#1{#2#3}{\mathrm{T}}$}
     \raise2\p@\vbox{\hbox{$#2#3$}}\kern-.35\wd0}}
\makeatother

\begin{document}

\title{Currents in a Tokamak}


\author{Robert W. Johnson}
\email[]{rob.johnson@gatech.edu}
\affiliation{Atlanta, GA 30238, USA}
%\homepage[]{http://www.frc.gatech.edu/Rob_Johnson.htm}
%\thanks{}


%\date{\today}
\date{January 8, 2008.}

\begin{abstract}
A self-consistent analysis of the currents in a tokamak yields the result that the parallel neutralizing current is given simply by the sum of currents commonly called the bootstrap current and the Pfirsch-Schl\"{u}ter current.  An expression is given relating the parallel and perpendicular pressure gradient induced species flow velocities to the total current.
\end{abstract}

%\keywords{}

\maketitle

%\section{}
%\subsection{}
%\subsubsection{}

\section{Introduction}
With a self-consistent analysis of the equilibrium currents in a tokamak, one finds that the parallel neutralizing current is given simply by the sum of what are commonly called the bootstrap current and the Pfirsch-Schl\"{u}ter current.  As Woods\cite{woodsbook-06} correctly points out, the standard expression for the bootstrap current density\cite{kessel-1221,wang-3319,houlberg-3230}, $\mbf{J}_{BS} \simeq (-f_{trap}/\Bpol)(\ddrp)\;\bhat$, has a nonzero divergence, at odds with the usual property of source-less currents, $\divr \mbf{J}=0$.  What Woods fails to mention is that the standard expression for the Pfirsch-Schl\"{u}ter current density\cite{dendybook-93,dinkbook-05}, $\mbf{J}_{PS} \simeq ( -2 \epsi \Cth / \Bpol^0)(\ddrp)\;\bhat$, suffers the same fate.  Together, however, they form the parallel neutralizing current density, $\mbf{J}_\para$, which is without divergence.

\section{Derivation}
We begin with the usual equations of plasma equilibrium:
\beq
\del p = \mbf{J} \times \mbf{B}\;,\; \mu_0 \mbf{J} = \curl \mbf{B}\;,\; \divr \mbf{B} = \divr \mbf{J} = \mbf{B} \cdot \del p = \mbf{J} \cdot \del p = 0\;,
\eeq
where the total pressure is the sum of the species' pressures, $p=\sum_s n_s \Tbar_s=n_e \Tbar_e+(n_i+n_z)\Tbar_i$, and $\mbf{J}$ is the total current density.  We know that a present-day tokamak has an ohmic current, $\mbf{J}_\Omega$, and that we will be considering discharges with a neutral beam driven current, $\mbf{J}_{NBI}$.  We work in a concentric circular flux-surface geometry $(r,\theta,\phi)$, Figure~1, related to cylindrical coordinates $(R,\phi,Z)$ by $R=R^0+r \Cth \equiv R^0(1+\epsi \Cth)\;,\; Z=-r \Sth\;,\; \phi=\phi$, where $\phihat$ is taken along the direction of the plasma current.  Various quantities also are defined in a coordinate system aligned to the magnetic field, $(x,y,z)=(r,\perp,\para)$, Figure~2, where $\hat{z}=\mbf{B}^0/B^0 \equiv \bhat=(0,b_\theta, b_\phi)$, and we assume toroidal symmetry, $\partial / \partial \phi \Rightarrow 0$.  The magnetic field takes the form $\mbf{B}=\mbf{B}^0 / (1+\epsi \Cth)=(0,\Bpol,\Btor)$ in this geometry.  Some useful operators and identities\cite{book-nrl-77} are:
\beq
\del f = \left(\frac{\partial f}{\partial r}\;,\; \frac{1}{r}\frac{\partial f}{\partial \theta}\;,\; 0\right)\;,\; \divr \mbf{F} = \frac{1}{rR} \left[\frac{\partial}{\partial r}r R F_r + \frac{\partial}{\partial \theta}R F_\theta\right]\;,
\eeq
\beq
\curl \mbf{F} = \left(\frac{1}{rR}\frac{\partial}{\partial \theta}R F_\phi\;,\; \frac{-1}{R}\frac{\partial}{\partial r}R F_\phi\;,\; \frac{1}{r} \left[\frac{\partial}{\partial r}r F_\theta - \frac{\partial}{\partial \theta}F_r\right] \right)\;,
\eeq
\beq
\curl \del f = \divr \curl \mbf{F} = 0\;,\; \divr(\mbf{A}\times\mbf{B})=\mbf{B}\cdot\curl\mbf{A}-\mbf{A}\cdot\curl\mbf{B}\;,\; \mbf{A}\cdot\mbf{B}\times\mbf{C}=\mbf{C}\cdot\mbf{A}\times\mbf{B}\; .
\eeq
We note that for nonzero $\mbf{J}\times\mbf{B}$, $\del p = (\ddrp)\; \rhat$.  The ohmic and NBI currents are solely in the toroidal direction, $\mbf{J}_\Omega+\mbf{J}_{NBI}=(J_\Omega+J_{NBI})\;\phihat$, as is the measured total current $\mbf{J}=J\;\phihat$, and satisfy $\divr(\mbf{J}_\Omega+\mbf{J}_{NBI})=0$.  The sum of the driven and flow currents is $\mbf{J}=\mbf{J}_\Omega+\mbf{J}_{NBI}+\mbf{J}_\perp+\mbf{J}_\para$, where $\mbf{J}_\perp$ is the diamagnetic current and $\mbf{J}_\para$ is the neutralizing current.  The divergence of the diamagnetic current is given by
\beq
\divr\mbf{J}_\perp = \divr \left(\frac{\mbf{B}\times\del p}{B^2}\right)=(1/B^2)\del p\cdot \curl\mbf{B} + \mbf{B}\cdot\del p \times \del 1/B^2 - \mbf{B}/B^2\cdot\curl\del p = 0\;,
\eeq
and we find that $J^\perp_\theta=p'\Btor/B^2$ and $J^\perp_\phi=-p'\Bpol/B^2$, for $p'=\ddrp$.  Turning our attention now to the parallel neutralizing current $\mbf{J}_\para$, we find that in order to cancel the poloidal current coming from $\mbf{J}_\perp$, we must have $J^\para_\theta=-J^\perp_\theta=-p'\Btor/B^2$ .  Then, by similar triangles\cite{euclid-56}, Figure~3, we have
\beq
J^\para_\phi=\frac{\Btor}{\Bpol}J^\para_\theta=\frac{-p'}{\Bpol^0}\left(\frac{\Btor^0}{B^0}\right)^2(1+\epsi \Cth) \equiv J^{BS}_\phi+J^{PS}_\phi\;,
\eeq
where we identify the first term as the bootstrap current and the second term as the Pfirsch-Schl\"{u}ter current.  That $\divr \mbf{J}_\para$ equals zero is given by construction upon noticing that the divergence picks up only the poloidal component.  So far, no mention has been made of ``passing'' versus ``trapped'' particles, but we note that only passing particles may participate in flowing currents,
\beq
J=f_{pass} \left[\sigma_\phi E_\phi + \sum_s n_s e_s \left(u^\perp_\phi+u^\para_\phi\right)_s\right]+J_{NBI}\;,
\eeq
for toroidal conductivity $\sigma_\phi=|b_\phi\sigma_\para-b_\theta\sigma_\perp|$ and pressure gradient induced species flows $\mbf{u}^\perp_s$ and $\mbf{u}^\para_s$.  Thus, one may determine the $E_\phi$ profile given the total current, pressure, and magnetic field profiles.

\section{Conclusions}
That the bootstrap current and the Pfirsch-Schl\"{u}ter current may combine to form a valid, source-less current density should not be a surprise, as the calculation by Ross\cite{ross-473} has indicated for some time.  The magnitude of the diamagnetic and neutralizing currents' contribution to the total plasma current increases with the pressure gradient, suggesting that non-inductive current drive may suffice for sufficiently steep plasma pressure profiles.

%\begin{acknowledgments}
%\end{acknowledgments}

% Specify following sections are appendices. Use \appendix*  if there
%\appendix*

\newpage

%\bibliography{../plasma.bib}

\begin{thebibliography}{9}
\expandafter\ifx\csname natexlab\endcsname\relax\def\natexlab#1{#1}\fi
\expandafter\ifx\csname bibnamefont\endcsname\relax
  \def\bibnamefont#1{#1}\fi
\expandafter\ifx\csname bibfnamefont\endcsname\relax
  \def\bibfnamefont#1{#1}\fi
\expandafter\ifx\csname citenamefont\endcsname\relax
  \def\citenamefont#1{#1}\fi
\expandafter\ifx\csname url\endcsname\relax
  \def\url#1{\texttt{#1}}\fi
\expandafter\ifx\csname urlprefix\endcsname\relax\def\urlprefix{URL }\fi
\providecommand{\bibinfo}[2]{#2}
\providecommand{\eprint}[2][]{\url{#2}}

\bibitem[{\citenamefont{{Woods}}(2006)}]{woodsbook-06}
\bibinfo{author}{\bibfnamefont{L.~C.} \bibnamefont{{Woods}}},
  \emph{\bibinfo{title}{Theory of Tokamak Transport: New Aspects for Nuclear
  Fusion Reactor Design}} (\bibinfo{publisher}{Wiley-VCH},
  \bibinfo{year}{2006}).

\bibitem[{\citenamefont{{Kessel}}(1994)}]{kessel-1221}
\bibinfo{author}{\bibfnamefont{C.~E.} \bibnamefont{{Kessel}}},
  \bibinfo{journal}{Nuclear Fusion} \textbf{\bibinfo{volume}{34}},
  \bibinfo{pages}{1221} (\bibinfo{year}{1994}),
  \urlprefix\url{http://stacks.iop.org/0029-5515/34/1221}.

\bibitem[{\citenamefont{Wang}(1998)}]{wang-3319}
\bibinfo{author}{\bibfnamefont{S.}~\bibnamefont{Wang}},
  \bibinfo{journal}{Physics of Plasmas} \textbf{\bibinfo{volume}{5}},
  \bibinfo{pages}{3319} (\bibinfo{year}{1998}),
  \urlprefix\url{http://link.aip.org/link/?PHP/5/3319/1}.

\bibitem[{\citenamefont{Houlberg et~al.}(1997)\citenamefont{Houlberg, Shaing,
  Hirshman, and Zarnstorff}}]{houlberg-3230}
\bibinfo{author}{\bibfnamefont{W.~A.} \bibnamefont{Houlberg}},
  \bibinfo{author}{\bibfnamefont{K.~C.} \bibnamefont{Shaing}},
  \bibinfo{author}{\bibfnamefont{S.~P.} \bibnamefont{Hirshman}},
  \bibnamefont{and} \bibinfo{author}{\bibfnamefont{M.~C.}
  \bibnamefont{Zarnstorff}}, \bibinfo{journal}{Physics of Plasmas}
  \textbf{\bibinfo{volume}{4}}, \bibinfo{pages}{3230} (\bibinfo{year}{1997}),
  \urlprefix\url{http://link.aip.org/link/?PHP/4/3230/1}.

\bibitem[{\citenamefont{{Dendy}}(1993)}]{dendybook-93}
\bibinfo{author}{\bibfnamefont{R.}~\bibnamefont{{Dendy}}},
  \emph{\bibinfo{title}{Plasma Physics: an Introductory Course}}
  (\bibinfo{publisher}{Cambridge University Press},
  \bibinfo{address}{Cambridge, England}, \bibinfo{year}{1993}).

\bibitem[{\citenamefont{Dinklage et~al.}(2005)\citenamefont{Dinklage, Klinger,
  Marx, and Schweikhard}}]{dinkbook-05}
\bibinfo{editor}{\bibfnamefont{A.}~\bibnamefont{Dinklage}},
  \bibinfo{editor}{\bibfnamefont{T.}~\bibnamefont{Klinger}},
  \bibinfo{editor}{\bibfnamefont{G.}~\bibnamefont{Marx}}, \bibnamefont{and}
  \bibinfo{editor}{\bibfnamefont{L.}~\bibnamefont{Schweikhard}}, eds.,
  \emph{\bibinfo{title}{Plasma Physics: Confinement, Transport and Collective
  Effects}} (\bibinfo{publisher}{Springer}, \bibinfo{year}{2005}).

\bibitem[{\citenamefont{{Book}}(1977)}]{book-nrl-77}
\bibinfo{author}{\bibfnamefont{D.~L.} \bibnamefont{{Book}}},
  \bibinfo{type}{Tech. Rep.} \bibinfo{number}{3332},
  \bibinfo{institution}{Naval Research Laboratory} (\bibinfo{year}{1977}).

\bibitem[{\citenamefont{Heath and Euclid}(1956)}]{euclid-56}
\bibinfo{author}{\bibfnamefont{T.~L.} \bibnamefont{Heath}} \bibnamefont{and}
  \bibinfo{author}{\bibnamefont{Euclid}}, \emph{\bibinfo{title}{The Thirteen
  Books of Euclid's Elements, Books 1 and 2}} (\bibinfo{publisher}{Dover
  Publications, Incorporated}, \bibinfo{year}{1956}), ISBN
  \bibinfo{isbn}{0486600882}.

\bibitem[{\citenamefont{{Ross}}(1995)}]{ross-473}
\bibinfo{author}{\bibfnamefont{D.~W.} \bibnamefont{{Ross}}},
  \bibinfo{type}{Tech. Rep.} \bibinfo{number}{FRCR-473},
  \bibinfo{institution}{Fusion Research Center, University of Texas (Austin)}
  (\bibinfo{year}{1995}).

\end{thebibliography}

% If in two-column mode, this environment will change to single-column
% format so that long equations can be displayed. Use
%\begin{widetext}
%\end{widetext}

% figures should be put into the text as floats.
% Use the graphics or graphicx packages (distributed with LaTeX2e)
% and the \includegraphics macro defined in those packages.
% See the LaTeX Graphics Companion by Michel Goosens, Sebastian Rahtz,
% entire page. There is no need to do explicit centering.

\newpage

\begin{figure}%
\includegraphics[scale=.5]{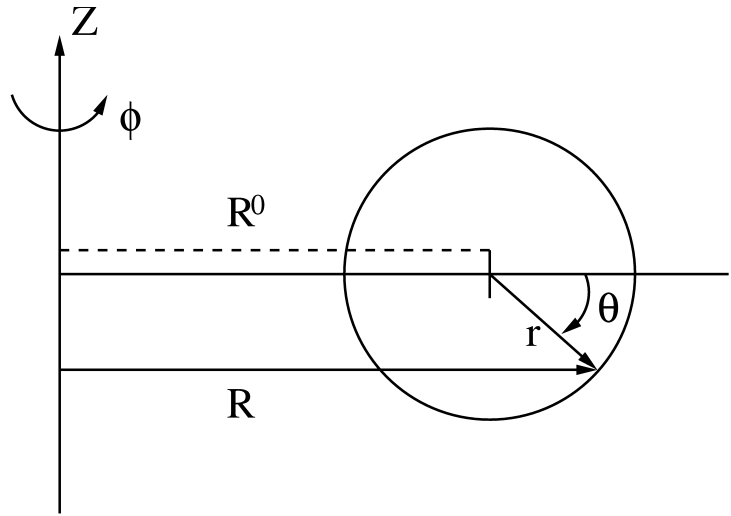}
\caption{Toroidal Coordinates.}
\end{figure}

\begin{figure}%
\includegraphics[scale=.5]{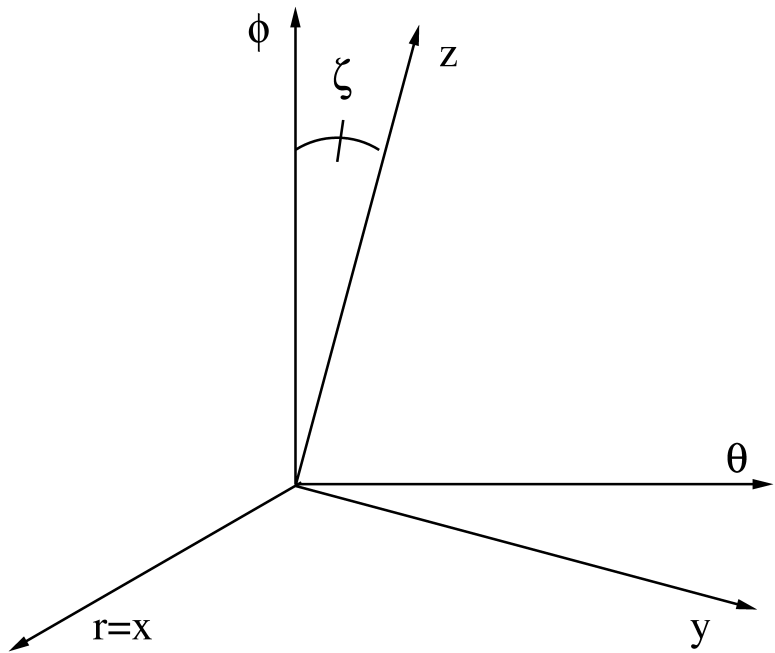}
\caption{Magnetic Coordinates.}
\end{figure}

\begin{figure}%
\includegraphics[scale=.5]{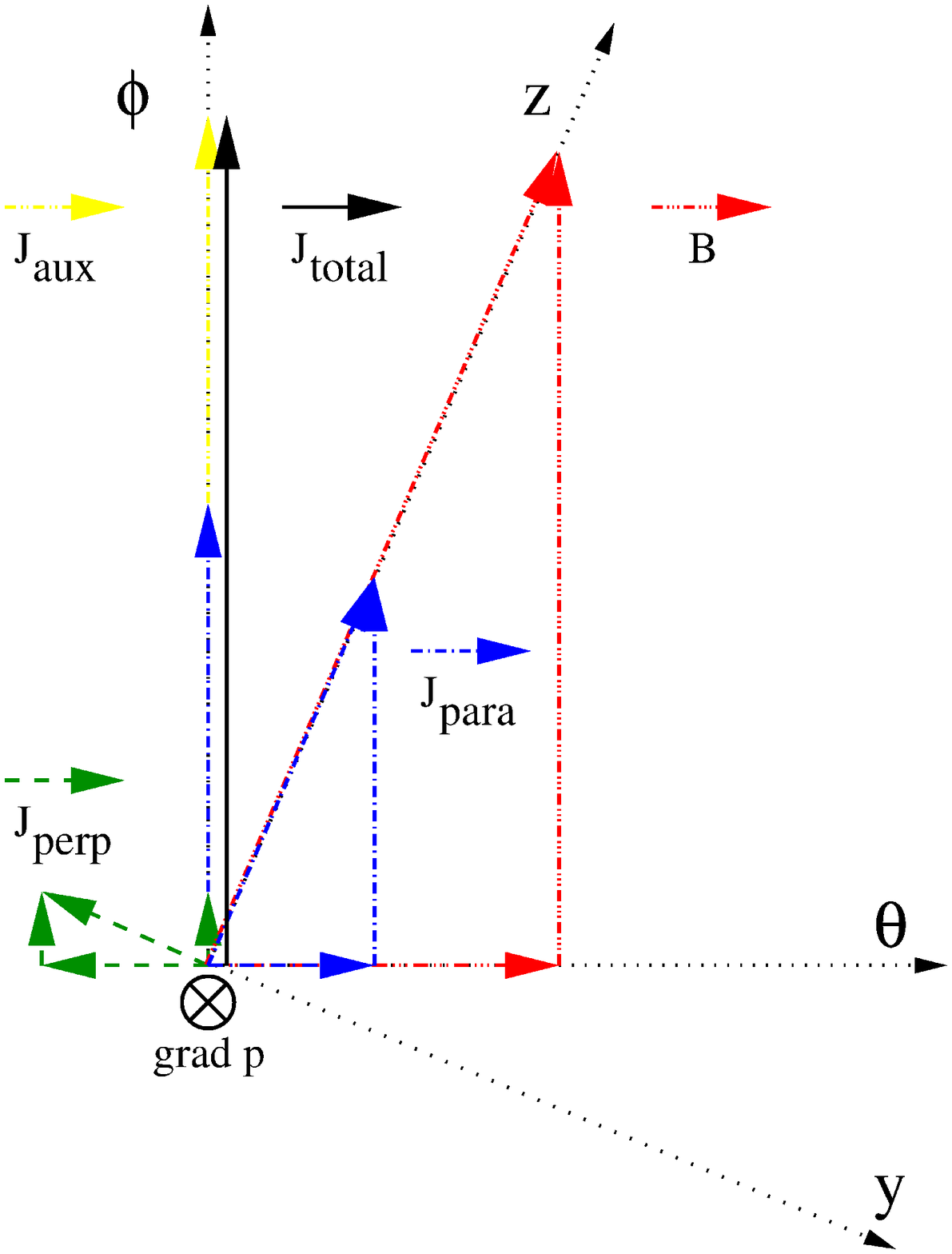}
\caption{(Color online.)  Magnetic Field and the Diamagnetic, Neutralizing, Driven, and Total Currents.}
\end{figure}

\end{document}